\documentstyle[aps,pra,epsfig]{revtex}

\begin{document}

\draft

\title{Deformed versus 
undeformed cat states encoding qubit}

\author{Stefano Mancini$^{1,2}$
and Vladimir I. Man'ko$^{3}$
}

\address{
$^{1}$INFM, Dipartimento di Fisica,
Universit\`a di Milano,
Via Celoria 16, I-20133 Milano, Italy
\\
$^{2}$Dipartimento di Matematica e Fisica,
Universit\`a di Camerino,
I-62032 Camerino, Italy
\\
$^{3}$P. N. Lebedev Physical Institute, Leninskii Prospekt 53,
Moscow 117924, Russia}

\maketitle

\widetext

\begin{abstract}
We study the possibility of exploiting superpositions of coherent 
states to encode qubit. A comparison between the use of deformed 
and undeformed bosonic algebra is made in connection 
with the amplitude damping errors.
\end{abstract}

\section{Introduction}

Controlling quantum coherence is one of the most 
fundamental issues in 
modern information processing \cite{QI}. 
The most popular solution in the
field of quantum information are quantum error correction codes
\cite{QEC} and error avoiding codes \cite{EAC}, both based on 
encoding the state into 
carefully selected subspaces of a larger Hilbert space
involving ancillary systems. 
The main limitation of these strategies
for combating decoherence is the 
large amount of extra space resources required \cite{STEANE}; 
in particular,
if fault tolerant error correction is also considered, 
the number of 
ancillary qubits enormously increases.
For this reason, other alternative 
approaches which do not require any ancillary resources have been 
pursued \cite{KICKS}.

In quantum information theory logical states are encoded as two 
orthogonal pure states \cite{QI}.  
The simplest example is provided by
a single two-level system. However, there is no fundamental
reasons to restrict oneself to physical system with two dimensional
Hilbert space for the encoding. It may be more convenient to
encode logical states  as a superposition over a large 
number of basis states.

On the other hand, 
while the coupling with the environment is fixed, we are
free to choose how we encode the qubits, hence the choice of the 
basis for the logical encoding may change the error introduced.

In this paper we study the qubit encoding in the superposition
of coherent states of a bosonic mode. This latter will be
considered from the generic approach of deformed algebra 
\cite{BIE}, showing that deformation can be profitably used to 
reduce amplitude damping errors.

\section{Cat states encoding qubit}

Let us introduce a $f$-coherent state 
defined \cite{MMZS} as the eigenstate of
the annihilation operator of a $f$-deformed
bosonic field $A=a\sqrt{f(a^{\dag}a)}$,
where $f$ is an operator-valued function of the number operator
(here it is assumed Hermitian and real) and 
$a$ being the annihilation operator of the undeformed field.
In general, $f$ can be made dependent on continuous parameters, 
in such a way that, for given particular values, the usual algebra 
is recovered.  
The $f$-coherent state can be written as  
\begin{equation}\label{FCOH}
|\zeta, f\rangle={\cal N}\sum_{n=0}^{\infty}
\frac{\zeta^n}{\sqrt{[n]_f!}}|n\rangle\,,
\quad
{\cal N}=\left[\exp_f(\zeta^2)\right]^{-1/2}\,,
\end{equation}
where we have considered the amplitude
$\zeta\in {\bf R}$, and we have introduced
\begin{eqnarray}\label{EXPF}
&&\exp_f(x)\equiv\sum_{n=0}^{\infty}\frac{x^n}{[n]_f!}\,,
\\
&&[n]_f!\equiv\left[nf(n)\right]\times
\left[(n-1)f(n-1)\right]\times\ldots
\left[2 \, f(2)\right]
\times\left[f(1)\right]
\times\left[f(0)\right]\,.
\end{eqnarray}
The function $\exp_{f}$ is a deformed version of 
the usual exponential function.
They become coincident when $f$ is the identity.
Notice that $\exp_f(x)\exp_f(y)\ne \exp_f(x+y)$,
i.e. we have a non-extensive exponential 
which can be found in many physical problems
\cite{TSALLIS}.

Let us now consider the superpositions \cite{PLA}
\begin{eqnarray}\label{FCAT}
|\Phi_+\rangle&=&{\cal N}_+\left(|\zeta,f\rangle
+|-\zeta,f\rangle\right)\,,
\quad
{\cal N}_+=\left[2+2{\cal N}^2
\exp_f(-\zeta^2)
\right]^{-1/2}\,,
\\
|\Phi_-\rangle&=&{\cal N}_-\left(|\zeta,f\rangle
-|-\zeta,f\rangle\right)\,,
\quad
{\cal N}_-=\left[2-2{\cal N}^2
\exp_f(-\zeta^2)
\right]^{-1/2}\,.
\end{eqnarray}
These states represent a generalization
of the well known even and odd cat states \cite{DMM},
and reduce to them whenever $f\to {\bf 1}$. 

Since $|\Phi_{+}\rangle$ and 
$|\Phi_{-}\rangle$ are orthogonal, 
we are led to the following logical encoding 
for a single qubit \cite{ICSSUR}
\begin{eqnarray}\label{ENCODE}
|\;\overline{0}\;\rangle&\equiv& |\Phi_+\rangle=
{\cal N}_+\left(|\zeta,f\rangle
+|-\zeta,f\rangle\right)\,,
\\
|\;\overline{1}\;\rangle&\equiv& |\Phi_-\rangle=
{\cal N}_-\left(|\zeta,f\rangle
-|-\zeta,f\rangle\right)\,.
\end{eqnarray}
In case of no deformation this reduces to the encoding
procedure proposed in Ref.\cite{COC}.
In such a case ${\cal N}_{+}$ and ${\cal N}_{-}$
tend to become equal as soon as $|\zeta|>1$.
Instead, in the general case, 
their difference drastically depends on 
the field deformation.
This can be evaluated by introducing the parameter
\begin{equation}
\Delta=\frac{|{\cal N}_{+}-{\cal N}_{-}|}
{{\rm min}[{\cal N}_{+},{\cal N}_{-}]}\,,
\end{equation}
which represents the relative error done 
by assuming ${\cal N}_{+}={\cal N}_{-}$.
$\Delta$ plays an important role in the qubit operations
as we shall see.

A further parameter which characterizes the cat states 
(hence our encoded qubit) in equation (\ref{ENCODE}) is the
the separation between the two superposed states \cite{DOD}.
This distance can be written as  
\begin{eqnarray}\label{FDIST}
d&\equiv&\langle\zeta,f|\left(a+a^{\dag}\right)|\zeta,f\rangle
\\
&=&{\cal N}^2 \, 
\sum_{n=0}^{\infty}\frac{\zeta^n}{\sqrt{[n]_f!}}
\left\{
\frac{\sqrt{n}\,\zeta^{(n-1)}}{\sqrt{[n-1]_f!}} 
+\frac{\sqrt{n+1}\,\zeta^{(n+1)}}{\sqrt{[n+1]_f!}}
\right\}\,.
\end{eqnarray}
For the case $f\to{\bf 1}$, 
we know \cite{PHOENIX} that the 
amplitude damping take places on a time scale 
inversely proportional to $d$.
Thus, for a given $\zeta$, the possibility to change $d$
through a suitable algebraic deformation 
(see, e.g. Ref.\cite{EPL2}) would
be very important to prevent errors on the
encoded qubit.

Among the infinite possible choices of $f$ we are going to
consider 
\begin{equation}\label{FLAG}
f(n)=\frac{L^{1}_{n}(\xi^{2})}{(n+1)L^{0}_{n}(\xi^{2})}\,,
\quad 
\xi\in{\bf R}\,,
\end{equation}
which we name $L$-deformation,
since $L^m_n$ indicates the associate 
Laguerre polynomial. 
Such type of deformation arises
in ion trapped systems (e.g.,
when an ion is bichromatically driven far from
the Lamb-Dicke regime) \cite{VOGEL}, 
then it could be accessible just in systems actually 
used for experimental quantum information
(see e.g., \cite{KIELP}).

\begin{figure}[t]
\centerline{\epsfig{figure=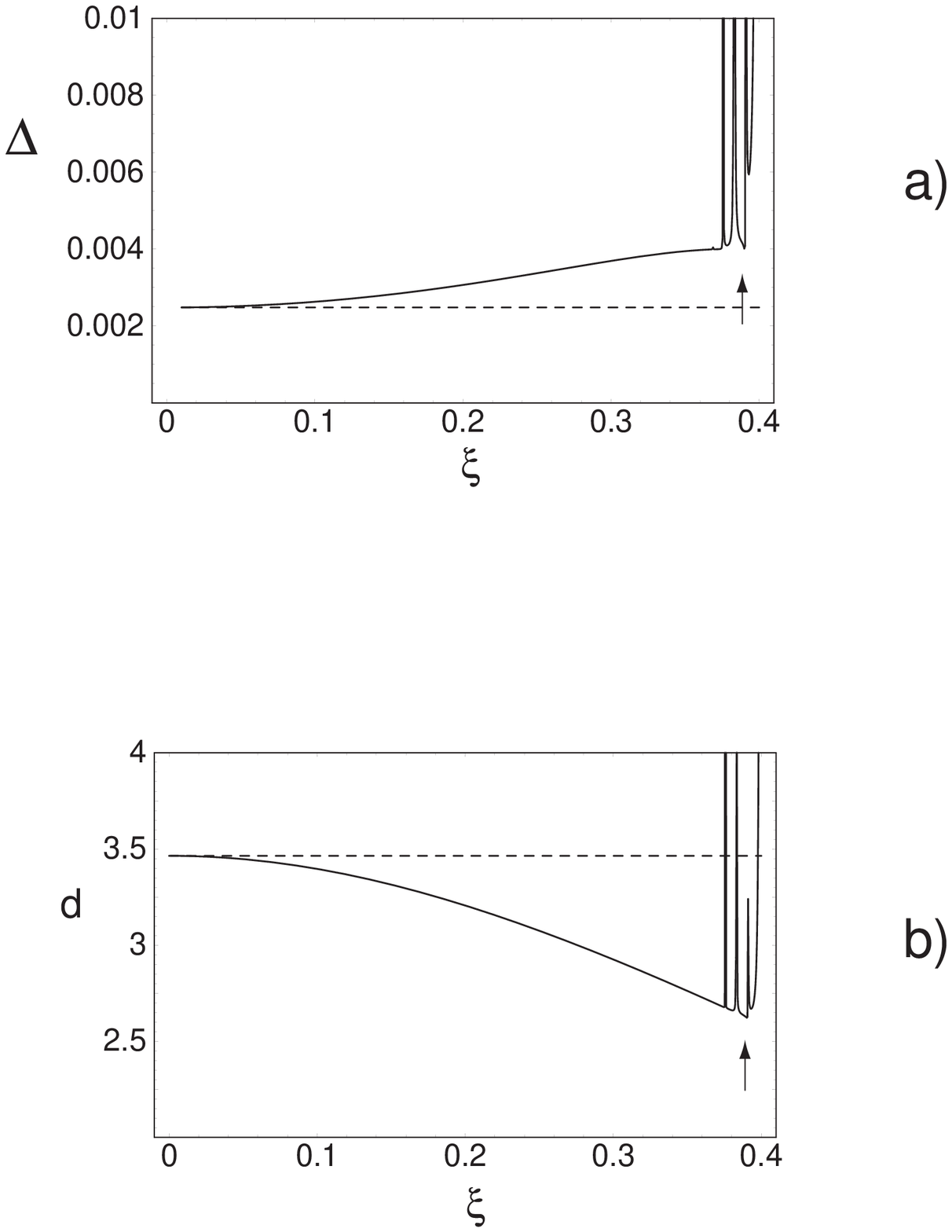,width=3.0in}}
\caption{
Quantity $\Delta$ versus parameter deformation $\xi$ (a).
State separation $d$ versus parameter deformation $\xi$ (b).
In both plots $\zeta^{2}=3$, the dashed line represents the 
non-deformed case and the solid line the $L$-deformation.
Small arrows indicate the value of $\xi$ used in figure \ref{fig2}.
}
\label{fig1}
\end{figure}

In figure \ref{fig1}a we show the behaviour of $\Delta$
as function of parameter $\xi$, while
in figure \ref{fig1}b we have plotted the distance $d$
as function of parameter $\xi$.
From these figures results
the possibility to have a distance $d$
smaller than the undeformed case still maintaining
$\Delta\approx 0$ when $|\zeta|>1$.

\section{Amplitude damping errors}

The amplitude damping errors on the qubit (\ref{ENCODE}) can be 
ascribed to
a dissipative interaction with an environment \cite{ZUR}.
This can be described (in interaction picture)
by the following master equation of the Lindblad form \cite{GAR}
\begin{equation}\label{ME}
\dot{\rho}=\gamma\, a\rho a^{\dag}
-\frac{\gamma}{2} \left\{a^{\dag}a,\rho\right\}\,,
\end{equation}
where $\gamma$ is the damping rate,
and we have set the bath temperature equal to zero.
The decoherence effect on the state
$\rho(0)=|\Psi_{\pm}\rangle\langle\Psi_{\pm}|$
can be described in the following way \cite{GJM}
\begin{equation}\label{RHOT}
\rho(t)=\sum_{k=0}^{\infty}
\Upsilon_k(t)\rho(0)\Upsilon^{\dag}_k(t)\,,
\end{equation}
where
\begin{equation}\label{UPS}
\Upsilon_k(t)=\sum_{n=k}^{\infty}
\sqrt{
\left(
\begin{array}{c}
n\\k
\end{array}
\right)}\;
\left[\eta(t)\right]^{(n-k)/2}\,\left[1-\eta(t)\right]^{k/2} \,  
|n-k\rangle\langle n|\,,
\end{equation}
with $\eta(t)=e^{-\gamma t}$.

In Ref.\cite{EPL2} the survival of the quantum coherence 
in deformed cat states has been shown.
Here,
the robustness of deformed cat states (qubit) 
against dissipative 
decoherence can be seen by considering the fidelity
\begin{equation}\label{FIDE}
{\cal F}(t)={\rm Tr}\left\{\rho(t)\rho(0)\right\}\,,
\end{equation}
which tell us to what extent the evolved state remains 
faithful to the initial one.
Starting from $\rho(0)=|\Phi_{\pm}\rangle\langle\Phi_{\pm}|$
we get
\begin{eqnarray}
{\cal F}_{\pm}(t)&=&
{\cal N}^4{\cal N}_{\pm}^4\sum_{k=0}^{\infty}
\;
\sum_{n,m=k}^{\infty}
\sqrt{
\left(
\begin{array}{c}
n
\\
k
\end{array}
\right)
\left(
\begin{array}{c}
m
\\
k
\end{array}
\right)}
\left[\eta(t)\right]^{(n+m)/2-k}
\left[1-\eta(t)\right]^{k}
\nonumber\\
&\times&
\frac{(\zeta)^n\pm(-\zeta)^n}
{\sqrt{[n]_f!}}
\times
\frac{(\zeta)^m\pm(-\zeta)^m}
{\sqrt{[m]_f!}}
\nonumber\\
&\times&
\frac{(\zeta)^{n-k}\pm(-\zeta)^{n-k}}
{\sqrt{[n-k]_f!}}
\times
\frac{(\zeta)^{m-k}\pm(-\zeta)^{m-k}}
{\sqrt{[m-k]_f!}}\,.
\end{eqnarray}

\begin{figure}[t]
\centerline{\epsfig{figure=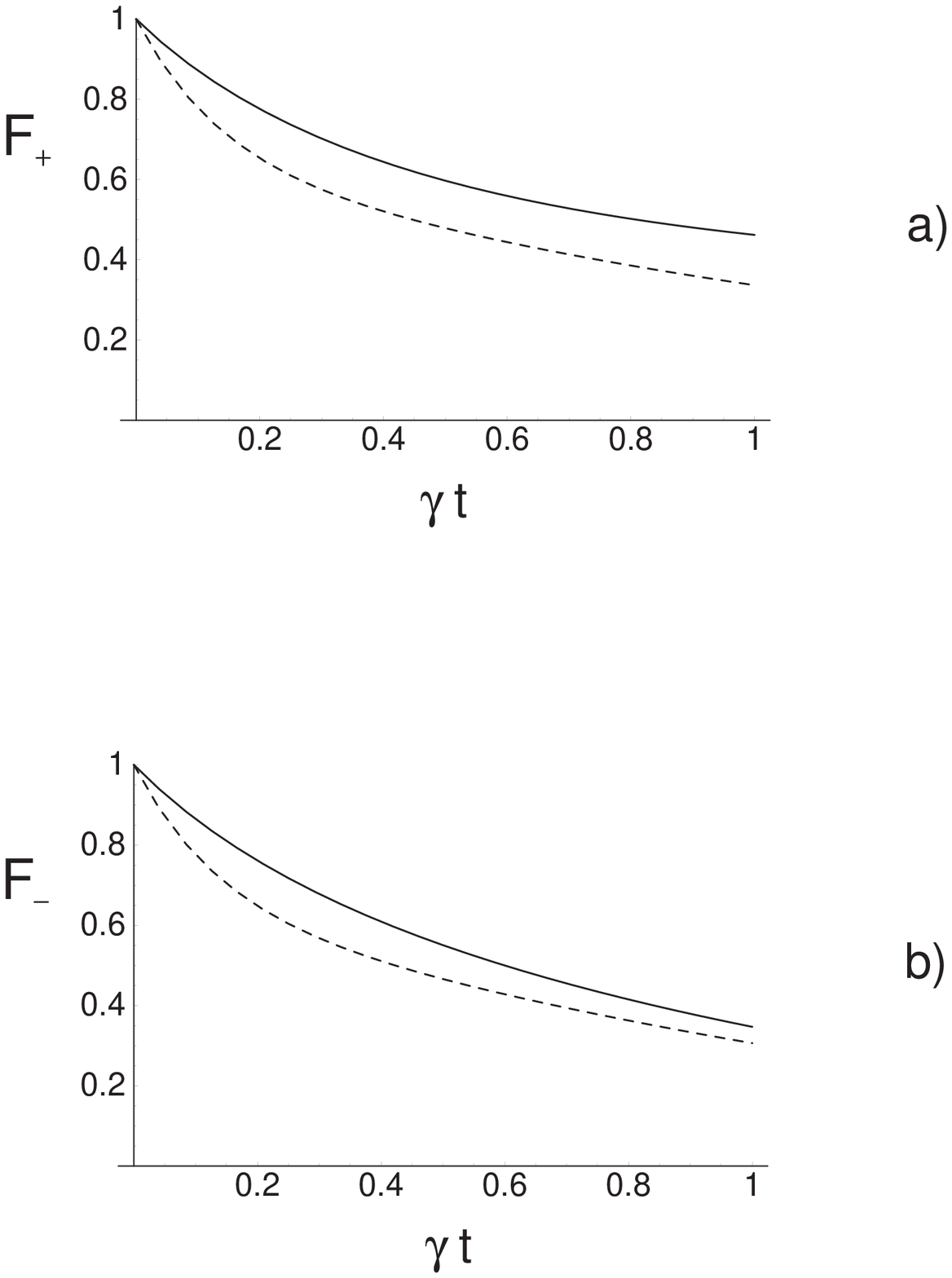,width=3.0in}}
\caption{
Fidelity ${\cal F}_{+}$ (a) and ${\cal F}_{-}$ (b)
as function of dimensionless time $\gamma t$, 
for $\zeta^{2}=3$.
The dashed line represents the non-deformed case while 
the solid lines refer to $L$-deformed case.
The used value of $\xi$ corresponds to 
that indicated by the arrows in figure \ref{fig1}.}
\label{fig2}
\end{figure}

In Fig.\ref{fig2} we compare (in time) the fidelity 
of an undeformed 
cat state (dashed line) with that 
of a deformed one (solid line).
We see the possibility to improve the fidelity 
by introducing an algebraic deformation on the 
bosonic mode.
This is essentially due to the reduced effective distance
between the two superposed states and to the fact that 
the latter (once deformed) are no longer eigenstates
(nor near eigenstates) of the irreversible
operator appearing in Eq.(\ref{ME}) \cite{ZUR}. 
It is also worth noting
the asymmetry between ${\cal F}_{+}$ and ${\cal F}_{-}$
which shows that the state encoded on the even cat 
is more robust than that encoded on the odd. 
The used values of parameter $\xi$ guarantees that also
$\Delta\approx 0$.

Better results could be obtained by exploring other regions of
the parameter $\xi$, however, this requires noticeable computer 
resources.

\section{Logical operations}

A logical encoding is useless if we cannot implement one and 
two qubit operations on the encoded states.
We now show Hamiltonians suitable to perform the fundamental
logical operations by generalizing the arguments of Ref.\cite{COC}.

For what concern the single qubit rotation we can 
construct the Hamiltonian generating such transformation
by simply using a driving term, that is
\begin{equation}\label{HR}
H_{R}=\beta A^{\dag}+\beta^{*} A
=\beta\sqrt{f(a^{\dag}a)}a^{\dag}
+\beta^{*} a\sqrt{f(a^{\dag}a)}\,,
\end{equation}
with $\beta$ the complex driven amplitude.
The time evolution under such Hamiltonian,
for small value of $|\beta|t$,
can be described by using the split operator 
method\footnote{
This form of the split operator method is only accurate to 
first order in $|\beta|t$, since it ignores higher order terms
involving the commutator of $A$ and $A^{\dag}$.
However, this method is superior to expanding $U$
to first order in $|\beta|t$ because it evolves the state
unitarily. Higher-order split operator methods also exist
\cite{FLECK}.}
\begin{equation}
U\approx \exp\left[-i\beta t A^{\dag}\right]
\exp\left[-i\beta^* t A\right]\,.
\end{equation}
Then, for a sufficiently 
large value of  $\zeta$, for which it is also $\Delta\approx 0$,
we have
\begin{eqnarray}\label{U}
U|\;\overline{0}\;\rangle&\approx&
\cos\theta   |\;\overline{0}\;\rangle
-i\sin\theta  |\;\overline{1}\;\rangle\,,
\\
U|\;\overline{1}\;\rangle&\approx&
\sin\theta  |\;\overline{0}\;\rangle
-i\cos\theta |\;\overline{1}\;\rangle\,,
\end{eqnarray}
where $\theta=2\zeta\beta t$ and we assumed $\beta\in{\bf R}$.
Equation (\ref{U}) is equivalent to the Hadamard transform provided
to have $\theta =\pi/4$. Notice that the operation (\ref{U}), hence
Hadamard transform, is possible just when 
${\cal N}_{+}\simeq {\cal N}_{-}$. 

For what concern the two-qubit gate, the simplest way to 
realize a universal gate is to employ the following 
Hamiltonian
\begin{equation}
H_{CPS}=\chi g^{-1}(A^{\dag}A)g^{-1}(B^{\dag}B)
=\chi a^{\dag}ab^{\dag}b\,,
\end{equation}
where $b$, $b^{\dag}$ ($B$, $B^{\dag}$) 
are the undeformed (deformed) ladder operators
of the second mode.
Furthermore $g(x)=xf(x)$.

If the interaction time is such that $\chi t=\pi$, we have 
\begin{eqnarray}
e^{-iH_{CPS}t}|\;\overline{0}\;\rangle|\;\overline{0}\;\rangle
&=&|\;\overline{0}\;\rangle|\;\overline{0}\;\rangle\,,
\\
e^{-iH_{CPS}t}|\;\overline{1}\;\rangle|\;\overline{0}\;\rangle
&=&|\;\overline{1}\;\rangle|\;\overline{0}\;\rangle\,,
\\
e^{-iH_{CPS}t}|\;\overline{0}\;\rangle|\;\overline{1}\;\rangle
&=&|\;\overline{0}\;\rangle|\;\overline{1}\;\rangle\,,
\\
e^{-iH_{CPS}t}|\;\overline{1}\;\rangle|\;\overline{1}\;\rangle
&=&-|\;\overline{1}\;\rangle|\;\overline{1}\;\rangle\,,
\end{eqnarray}
The above two-qubit gate represents a conditional phase
shift, and can be easily understood by reminding that
$|\,\overline{0}\,\rangle$ contains only even bosonic
and $|\,\overline{1}\,\rangle$ only odd bosonic number.

Here, we do not deal the question of how to implement the above 
interactions, but we simply postulate their existence.
On the other hand, an arbitrary unitary 
transformation can be built up efficiently, to any desired
precision, by using elementary interactions 
\cite{LLOYD}, realizable in several systems \cite{QI}.
In particular, there already exist 
proposals to engineer any Hamiltonian for a trapped ion
\cite{MATOS}.

\section{Conclusion}

In conclusion, we have shown that quantum information
can be suitably encoded in cat states.
Moreover, the deformed version of such states offers
the possibility of reducing amplitude damping errors.
The introduction of algebraic field deformation 
to reduce the decoherence effects of qubit is
reminiscent of ``passive" strategy 
like ``error avoiding codes" \cite{EAC}.
However, in this case the whole Hilbert space is exploited
without waste of any degrees of freedom.

Beside trapped systems showing $L$-deformed states \cite{VOGEL}, 
one can look for other type of realistic 
deformations leading analogous effects. 
Potential candidates could be
Bose-Einstein condensates  
where the requirement of particle number conserving
leads to a modification of the field algebra \cite{BEC}.

Finally, the studied problem constitutes a building block
of a more general and intriguing problem 
concerning the group theoretical approach to field deformation
\cite{DEF} and decoherence \cite{SEMI} 
(e.g., find a suitable deformation such that ${\cal F}_{\pm}$
remains approximately one)
which we plan to deal in a future.


\begin{thebibliography}{99}

\bibitem{QI}
Nielsen M A and Chuang I L 2000
{\it Quantum Computation and Quantum Information}
(Cambridge: University Press)

\bibitem{QEC}
Shor P W 1995 {\it Phys Rev A} {\bf 52} 2493
\\
Steane A M 1995 {\it Proc R S A} {\bf 452} 2551
\\
Knill E and Laflamme R 1997 {\it Phys Rev A} {\bf 55} 900

\bibitem{EAC}
Zanardi P and Rasetti M 1997 {\it Phys Rev Lett} {\bf 79}, 3306
\\ 
Duan L M and Guo G C 1997 {\it Phys Rev Lett} {\bf 79} 1953
\\ 
Lidar D A, Chuang I L and Whaley K B 1998 {\it Phys Rev Lett} 
{\bf 81} 2594 

\bibitem{STEANE}
Steane A M 1999 {\it Nature (London)} {\bf 399} 124

\bibitem{KICKS}
Viola L and Lloyd S 1998 {\it Phys Rev A} {\bf 58} 2733 
\\
Vitali D and Tombesi P 1999 {\it Phys Rev A} {\bf 59} 4718 

\bibitem{BIE}
Biedenharn L C 1989 {\it J Phys A} {\bf 22} L873 
\\
Macfarlane A J 1989 {\it J Phys A }{\bf 22} 4581 

\bibitem{MMZS}
Man'ko V I, Marmo G, Zaccaria F and Sudarshan E C G 1997
{\it Phys Scr} {\bf 55} 528

\bibitem{TSALLIS}
Tsallis C 1998 {\it J. Stat. Phys.} {\bf 52} 479 
\\
Tsallis C 1999 {\it Braz. J. Phys.} {\bf 29} 1 

\bibitem{PLA}
Mancini S 1997 {\it Phys Lett A} {\bf 233} 291 
\\
Roy B and Roychoudhury R 1997
{\it Int. J. Theor. Phys.} {\bf 36} 1525 

\bibitem{DMM}
Dodonov V V, Malkin I A and Man'ko V I 1974
{\it Physica} {\bf 72} 597

\bibitem{ICSSUR}
Mancini S and Man'ko V I 
{\it Proc. 7th Int. Conf.  
Squeezed States and Uncertainty Relation (Boston)}
to appear

\bibitem{COC}
Cochrane P T, Milburn G J and Munro W J 1999
{\it Phys Rev A} {\bf 59} 2631  

\bibitem{DOD}
Dodonov V V, Man'ko O V, Man'ko V I and Wuensche A 1999
{\it Phys Scr} {\bf 59} 81 
\\
Dodonov V V, Man'ko O V, Man'ko V I and Wuensche A 2000
{\it J. Mod. Phys.} {\bf 47} 633 

\bibitem{PHOENIX}
Phoenix S J D 1990 {\it Phys Rev A} {\bf 41} 5132 
\\
Buzek V, Vidiella-Barranco A and Knight P L
1992 {\it Phys Rev A} {\bf 45} 6570
\\
Kim M S and Buzek V 1992
{\it Phys Rev A} {\bf 46} 4239 

\bibitem{EPL2}
Mancini S and Man'ko V I 2001
{\it Europhys. Lett.} {\bf 54} 586

\bibitem{KIELP}
Kielpinski D, Meyer V, Rowe M A, Sackett C A,
Itano W M, Monroe C and Wineland D J 2001
{\it Science} {\bf 291} 1013

\bibitem{VOGEL}
de Matos Filho R L and Vogel W 1996
{\it Phys Rev A} {\bf 54} 4560 
\\
Kis Z, Vogel W and Davidovich L 2001
{\it Phys Rev A} {\bf 64} 033401

\bibitem{ZUR}
Zurek W H 1981 {\it Phys Rev D} {\bf 24} 1516
\\
Zurek W H 1982 {\it Phys Rev D} {\bf 26} 1862 
\\
Zurek W H 1991 {\it Phys. Today} {\bf 44}(10) 36

\bibitem{GAR}
Gardiner C W 1991 {\it Quantum Noise}
(Berlin: Springer)

\bibitem{GJM}
Milburn G J 1999 {\it Proc. Fifth Summer School on Atomic,
Molecular and Optical Physics}
(Singapore: World Scientific) p 435

\bibitem{FLECK}
Fleck J A, Morris J R and Fleit M D 1976
{\it Appl. Phys.} {\bf 10} 129

\bibitem{LLOYD}
Lloyd S and Braunstein S 1999 {\it Phys Rev Lett} {\bf 82} 1784

\bibitem{MATOS}
de Matos Filho R L and Vogel W 1998
{\it Phys Rev A} {\bf 58} R1661

\bibitem{BEC}
Sun C P, Yu S and Gao Y B 1998
{\it Preprint} quant-ph/9809079
\\
Mancini S and Man'ko V I 1999 
{\it Phys Lett A} {\bf 259} 67

\bibitem{DEF}
Ge M L 1992
{\it Quantum Groups and Quantum Integrable Systems}
(Singapore: World Scientific)

\bibitem{SEMI}
Alicki R and Lendi K 1987
{\it Quantum Dynamical Semigroups and Aplications}
Lecture Notes in Physics N 286 (Berlin: Springer)

\end{thebibliography}
\end{document}